\documentclass[12pt,aps,prb,preprint]{revtex4}   
\usepackage{graphicx}   
\usepackage{amsfonts}
\usepackage{amsmath}    
\usepackage{amssymb}
\usepackage{bm}
\usepackage{bbm}
\usepackage{fancyhdr}
\usepackage{textcomp}
\usepackage{mathrsfs}
\usepackage{enumerate}
\begin{document}
\title{A unique pure mechanical system revealing dipole repulsion}
\author{June-Haak Ee}
 \affiliation{Department of Physics, Korea University,
 Seoul 136-713, Korea}
\author{Jungil Lee}
 \email{jungil@hep.anl.gov}   
 \affiliation{Department of Physics, Korea University,
 Seoul 136-713, Korea}
\date{\today}
\begin{abstract}
We study multiple elastic collisions of a block 
and a ball against a rigid wall in one dimension. The complete 
trajectory of the block is solved as an analytic function of time.
Near the turning point of the block the force carried
by the ball is proportional to $1/x^3$, where $x$ is the distance 
between the wall and the block, in the limit that the block is 
sufficiently heavier than the ball. 
This is a unique pure mechanical system that reveals
dipole-like repulsion.
\end{abstract}
\maketitle
\section{Introduction\label{sec:intro}}
An elastic collision is a useful mechanical problem with which
one can study energy, linear-momentum, and angular-momentum
conservation in an explicit way.\cite{general-physics}
While the final state is uniquely determined in a two-body elastic
collision in one dimension, there are multiple solutions if three
or more particles are involved. The reason is that the two constraints
from energy and momentum conservation are not enough to determine
the three or more final-state momenta.
For example, a unique solution for a Newton's cradle 
of three or more pendula is obtained only if any two neighboring pendula
are separated so that each collision involves only two pendula. 

Various studies on such chain collisions 
have been carried out for the instantaneous 
contact force,\cite{Kline:1960,Kerwin:1972,%
Huebner:1992,Murphy:1994,Erlichson:1996,Patricio:2004}
for the spring force,\cite{Chapman:1960,Flansburg:1979,%
Herrmann:1981,Hinch:1999} and for the contact Hertz 
force.\cite{Chapman:1960,Hinch:1999,Hessel:2006,Patricio:2004}
In addition, there are numerous studies of the collisions 
of aligned balls in one dimension.\cite{%
Lemon:1935,Sleator:1937,Mellen:1968,Harter:1971,Stroink:1983,%
Bruce:1990,Mellen:1995,Cross:2007,Muller:2011}
Harter investigated the velocity gain in
such a chain collision.\cite{Harter:1971} 
Patricio studied
the effect of the Hertz contact force in detail.\cite{Patricio:2004}
Redner considered the one-dimensional
collisions of two cannonballs with a table-tennis ball sandwiched
between them to derive
a simple relation between that elastic collision and a
corresponding billiard system.\cite{Redner:2004}

In this paper, we consider the multiple collisions of
two particles in one dimension against a rigid wall. As is stated
above, this problem is, in principle, exactly solvable and
we attempt to find the analytic solution to this problem.
Many people, who have played or watched table-tennis games, 
are familiar with this multiple collision. When the server is 
waiting for the defender, the server often bounces the ball 
against the table repeatedly with the racket. While the 
player presses down the ball with the racket, the speed of the
ball keeps increasing and one
can hear an interesting high-frequency sound near the turning
point of the racket.

Whelan \textit{et al.} studied a similar
case under gravity focusing on its chaotic behavior.\cite{Whelan:1990}
Because the main feature of the problem stems from the velocity
amplification of the light target against the heavy incident
particle, we ignore any external forces other than the contact
interaction between any two colliding objects.
We construct a simple model system consisting of a block
and a rigid wall that sandwich a ball. 
We derive the analytic expression for the complete trajectory of the
block as a function of time.
This new analytic solution reveals that, when the ball is sufficiently
lighter than the block, the ball mediates a force proportional to $1/x^3$
near the turning point of the block. Here, 
$x$ is the distance between the wall and the block. 
This emergent dipole interaction is
unique in a pure mechanical system.
In comparison with a previous study in Ref.~\cite{Sinai:1991} for a
case with many balls, our derivation of the $1/x^3$ rule is based on 
a complete analytic solution with a rigorous error analysis.

This paper is organized as follows. In Sec.~\ref{sec:problem}
we introduce our model, which simplifies the multiple collision
problem of the table-tennis ball. In Sec.~\ref{sec:solution} we
construct the simultaneous recurrence relation for the velocities
of the colliding particles in each collision.
Those for the position and time are
also given. We determine the complete trajectory of the block
as a function of time.
In Sec.~\ref{sec:continuum}, 
we investigate the dynamics of the system in the continuum limit,
where the block is sufficiently heavier than the ball and the
block is near the turning point.
Our conclusion is given in Sec.~\ref{sec:concl},
followed by the appendices, which contain
useful mathematical formulas and some details of calculations.
\section{\label{sec:problem}The model system}
In this section, we define our model system. We restrict ourselves
to elastic collision in one spatial dimension and 
ignore any external forces  other than instantaneous
contact forces.

\begin{figure}[ht]
\begin{center}
\vspace{-10ex}
\includegraphics[width=80mm]{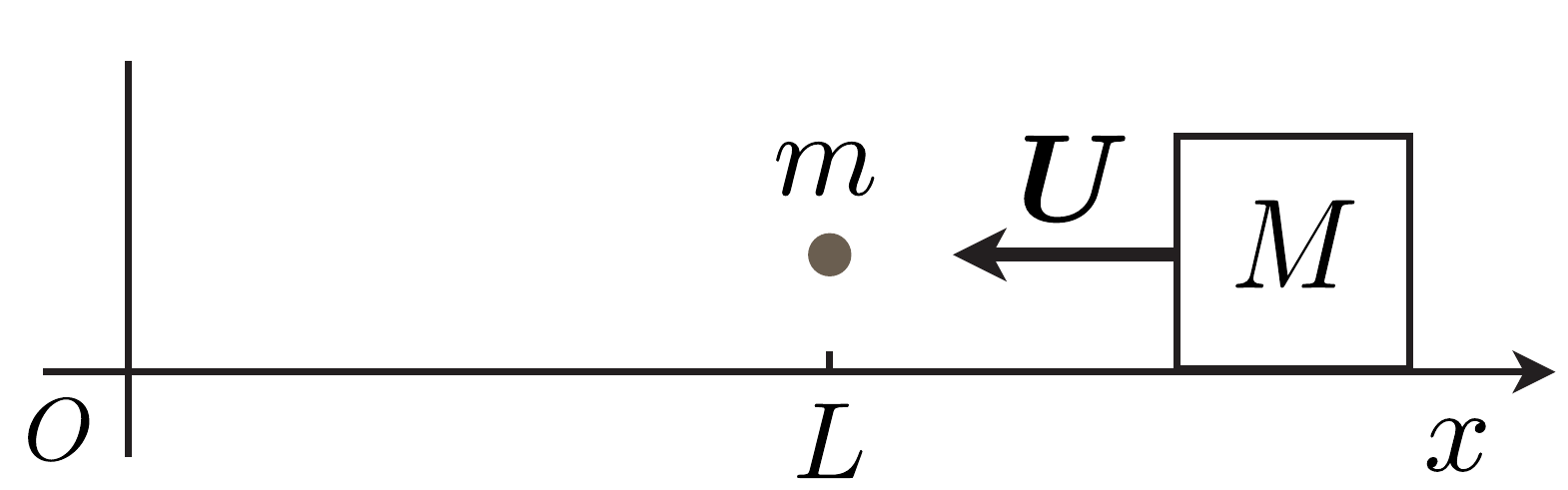}
\caption{\label{fig:initial}
The initial condition of the model system.}
\end{center}
\end{figure}

As shown in Fig.~\ref{fig:initial}, the model system consists of
a block with mass $M$, a ball with mass $m$,
and a rigid wall at $x\le 0$. The unit vector along the positive 
$x$ axis is given by $\bm{\hat{x}}$ and the mass ratio is defined by
\begin{equation}
\alpha={m}/{M}.
\end{equation}
At time $t=t_1=0$ the block hits the ball at $x=x_1=L$, where
initial velocities of the block and ball are
\begin{equation}
\label{IC}
U_0=-U\quad\textrm{and}\quad u_0=0,
\end{equation}
respectively. After the first collision, the ball bounces against
the wall and then they make the next collision.
We call $P_n(t_n,x_n)$ the $n$-th collision point 
between the ball and the block,
where $t_n$ and $x_n$ are the time and position.
The velocities of the block and the ball right after the $n$-th
collision are $U_n$ and $u_n$, respectively. The total number of
collisions $N$ is the smallest $n$ that satisfies the condition 
$U_n\ge|u_n|$.
We also define the time interval $\Delta t_n=t_{n+1}-t_n$ 
between $P_{n}$ and $P_{n+1}$. Useful relations involving
the computation of $t_n$ using $\Delta t_n$ are given in
Appendix \ref{app-tn}.

The elasticity of the collisions requires 
the conservation of energy as well as the linear momentum
in the $n$-th collision:
\begin{subequations}
\label{energy-momentum-n}
\begin{eqnarray}
MU_n-m u_{n} &=& M U_{n+1} + m u_{n+1},
\\
\frac{1}{2}MU_{n}^2+\frac{1}{2}m u_{n}^2
&=& \frac{1}{2}MU_{n+1}^2+\frac{1}{2}m u_{n+1}^2.
\end{eqnarray}
\end{subequations}
Note that the sign of $u_n$ is flipped in the first equation
because the ball has bounced against the wall.
With the initial conditions in Eq.~(\ref{IC}), one can solve
$U_n$ and $u_n$ recursively.
The first two pairs of the solutions are
\begin{subequations}
\label{Uu12}
\begin{eqnarray}
\label{Uu1}
\begin{pmatrix}
U_1\\{u}_1
\end{pmatrix}
&=&-\frac{U}{1+\alpha}
\begin{pmatrix}
1-\alpha\\
2
\end{pmatrix},
\\
\label{Uu2}
\begin{pmatrix}
U_2\\{u}_2
\end{pmatrix}
&=&-\frac{U}{(1+\alpha)^2}
\begin{pmatrix}
1-6\alpha+\alpha^2\\
4(1-\alpha)
\end{pmatrix}.
\end{eqnarray}
\end{subequations}
Based on Eqs.~(\ref{Uu12}), we classify the ranges
of the mass ratio $\alpha$ that determine the value of $N$.
\begin{itemize}
\item
If $\alpha\ge 3$, then $N=1$ because $U_1\ge -u_1$.
The block bounces back after the
first collision and they never collide again. At $\alpha = 3$,
$U_1=-u_1=\tfrac{1}{2}\,U$.
\item
If $1\le\alpha<3$, then $N=2$ because $U_1<-u_1$ and
$U_2> u_2\ge 0$.  At $\alpha=1$, $U_1=0$, $u_1=-U$,
$U_2=U$, and $u_2=0$.
 \item
If $\alpha< 1$, then $N\ge2$. Especially,
for an $\alpha< 5-2\sqrt{5}$, $N\ge 3$.
At $\alpha=5-2\sqrt{5}$, $U_2=-u_2$.
\end{itemize}
In the remainder of this paper,
we consider the case $0<\alpha<1$
that involves multiple collisions.
\section{\label{sec:solution}Analytic Solutions}
In this section, we determine the point $P_n(t_n,x_n)$
of the $n$-th collision between the ball and block, and the 
velocities $U_n$ and $u_n$ immediately after the collision.

\subsection{\label{sec:vel}Computation of $\bm{U_n}$ and $\bm{u_n}$}
We can reduce the recurrence relations for the velocities in 
Eq.~(\ref{energy-momentum-n}) into the form
\begin{equation}
\label{Uun}
\begin{pmatrix}
U_{n+1}\\u_{n+1}
\end{pmatrix}
=A
\begin{pmatrix}
U_{n}\\u_{n}
\end{pmatrix},
\end{equation}
where the $2\times 2$ matrix $A$ is defined by
\begin{equation}
\label{matrix-A}
A=\frac{1}{1+\alpha}
\begin{pmatrix}
1-\alpha&-2\alpha\\
2&1-\alpha
\end{pmatrix}.
\end{equation}
Applying Eq.~(\ref{Uun}) recursively, 
we determine $U_n$ and $u_n$:
\begin{equation}
\label{Uun-2}
\begin{pmatrix}
U_{n}\\u_{n}
\end{pmatrix}
=A^n
\begin{pmatrix}
U_0\\u_0
\end{pmatrix}.
\end{equation}
Although our initial conditions are given in Eq.~(\ref{IC}),
the relation (\ref{Uun-2}) is valid for any initial values
of $U_0$ and $u_0$.

The computation of $U_n$ and $u_n$ is rather tedious and
we summarize the calculation in Appendix~\ref{app-Uun}.
The results are
\begin{equation}
\label{sol-vnf}
\begin{pmatrix}
U_{n}\\u_{n}
\end{pmatrix}
=
-\,U
\begin{pmatrix}
\cos n\theta\\
\displaystyle\frac{1}{\sqrt{\alpha}}\sin n\theta
\end{pmatrix},
\end{equation}
where the parameter $\theta$ is related to the mass ratio $\alpha$ as
\begin{equation}
\label{theta-alpha}
\tan\tfrac{\theta}{2}=\sqrt{\alpha}.
\end{equation}
Let us interpret the results in Eq.~(\ref{sol-vnf}).
Because the total kinetic energy is the sum of quadratic
functions of $U_n$ and $u_n$, the conservation of the total
kinetic energy implies that $U_n$ and $u_n$ can be parametrized
as a cosine and a sine of a common phase. The phase increases as
a multiple of an elementary phase $\theta$ because the two eigenvalues
of the non-Hermitian matrix $A$ are a unimodulus complex number 
$e^{i\theta}$ and its complex conjugate.
According to these results, the block slows down to reach the
turning point around $n\theta\approx\tfrac{\pi}{2}$, where 
$|u_n|$ becomes the maximum.
The maximum value for $|u_n|$ is larger than the
initial speed $U$ of the block with the enhancement factor 
$1/\sqrt{\alpha}$, if $0<\alpha<1$.
From that moment to the region $n\theta\approx \pi$, the block
bounces back and the ball loses the kinetic
energy. If $n=N$, then the velocity $-u_n$ of the ball
after the bounce against the wall must satisfy the constraint
$U_n\ge -u_n$. Therefore, $N$ must be the smallest integer that
satisfies
\begin{equation}
\label{N-eq}
U_N+u_N\ge 0 \quad\textrm{and}\quad N >\frac{\pi}{2\theta}.
\end{equation}
\subsection{Computation of $\bm{x_n}$}
The recurrence relation of $x_n$ is derived in Eq.~(\ref{xn1xn-app})
of Appendix~\ref{app-tn}:
\begin{equation}
\label{xn1xn}
\frac{x_{n+1}}{x_n}=(u_n-U_n)/(u_n+U_n).
\end{equation}
By making use of Eqs.~(\ref{sol-vnf}) and 
(\ref{theta-alpha}) and the elementary
trigonometric identity
$\sin\alpha\cos\beta\pm\cos\alpha\sin\beta=\sin(\alpha\pm\beta)$,
we find that
\begin{equation}
\label{xn-qn}
\frac{x_{n+1}}{x_n}
=\frac
{\sin(n-\tfrac{1}{2})\theta}
{\sin(n+\tfrac{1}{2})\theta}.
\end{equation}
With the initial condition $x_1=L$, we find the analytic
expression for $x_n$ as
\begin{equation}
\label{xn-3}
x_n=
x_1\prod_{k=1}^{n-1}
\frac{x_{k+1}}{x_{k}}
=
\frac{L\sin\frac{\theta}{2}}
     {\sin(n-\tfrac{1}{2})\theta}.
\end{equation}
The analytic expression for $x_n$ is new.
From this compact expression, it is easy to see
that $x_n$ decreases to reach the minimum value $x_{\rm min}$ near
$n\approx \tfrac{1}{2}(\tfrac{\pi}{\theta}-1)$, where the denominator
reaches its maximum. The turning point $x_{\rm min}$
is reached at $n=N_1$ and the last collision is
made at $x_{N}$.

\subsection{Computation of $\bm{N_1}$ and $\bm{N}$}
From Eq.~(\ref{xn-3}), we can determine that $N_1$ is
\begin{equation}
\label{NN1}
N_1=\left[\frac{\pi}{2\theta}\right]_+,
\end{equation}
where $[x]_+$ is the smallest integer 
that is greater than or equal to $x$.
For a small $\alpha$, $\theta \approx 2\sqrt\alpha$ and therefore
$N_1\approx[\frac{\pi}{4\sqrt{\alpha}}]_+$. 
In analogy to the derivation of Eq.~(\ref{xn-3}),
we can simplify $U_n+u_n$ as
\begin{equation}
\label{Uun-ans}
U_n+u_n = 
-\frac{U\sin (n+\tfrac{1}{2})\theta}
       {\sin\tfrac{\theta}{2}}.
\end{equation}
Substituting Eq.~(\ref{Uun-ans}) into Eq.~(\ref{N-eq}),
we find that
\begin{equation}
\label{NN}
N=\bigg[\frac{\pi}{\theta}-\frac{1}{2}\bigg]_+
\approx[\frac{\pi}{2\sqrt{\alpha}}-\frac{1}{2}]_+,
\end{equation}
where the final expression is valid for small $\alpha$.
This result is consistent with Eq.~(6) of Ref.~\cite{Redner:2004}
with the uncertainty $\pm 1$. The quantity $N_{\rm{max}}$ 
of that reference is approximately equal to $2N$ because, 
in that reference, the collision between the ball and the wall is also counted.

\begin{figure}[ht]
\begin{center}
\includegraphics[width=80mm]{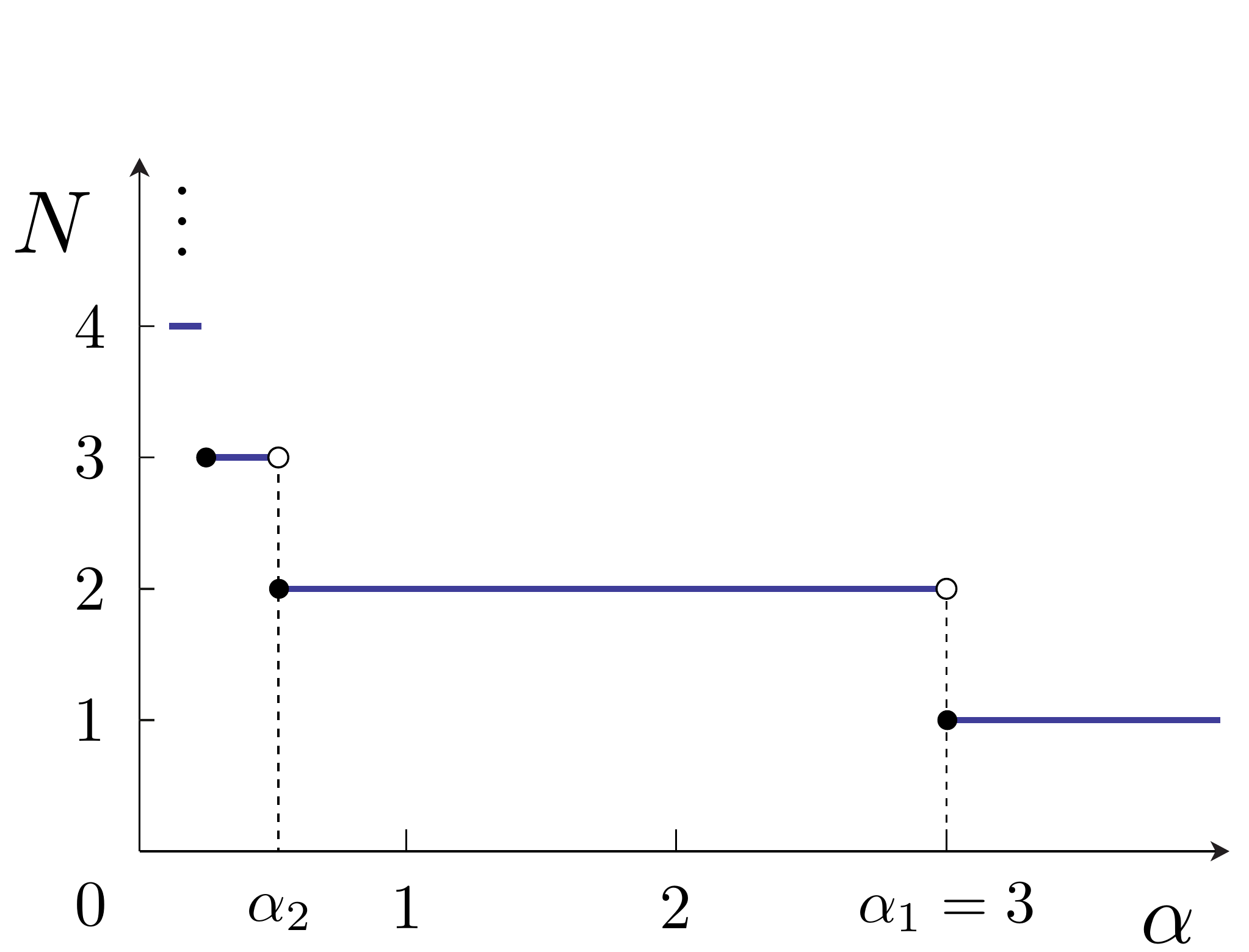}\\
\caption{\label{fig:N}
The total number of collision $N$ as a function of the
mass ratio $\alpha$. $\alpha_1=3$ and $\alpha_2=5-2\sqrt{5}$
are the minimum values of $\alpha$ to have $N=1$ and $N=2$,
respectively. As $\alpha \to 0^+$, 
$N\to \big[\tfrac{\pi}{2\sqrt{\alpha}}-\frac{1}{2}\big]_+$
which diverges to infinity.}
\end{center}
\end{figure}

In Fig.~\ref{fig:N}, we show the total number of collisions
$N$ as a function of $\alpha$.
The results in the figure are consistent with classification given
at the end of Sec.~\ref{sec:problem}. 
The discontinuities of $N$ are 
at $\alpha=\alpha_i$, where $\alpha_i$ is the minimum value
of $\alpha$ that satisfies $N=i$. The first two values of $\alpha_i$
are $\alpha_1=3$ and $\alpha_2=5-2\sqrt{5}$.
\subsection{Computation of $\bm{t_n}$}
We have determined $U_n$, $u_n$, and $x_n$ 
for all $n=1,\,\cdots,\, N$.
In this section, we use these results to compute the collision 
times $t_n$. The computation of the time interval $\Delta t_n$
is given in Appendix~\ref{app-dtn}. By making use of Eq.~(\ref{delta-tn}),
we sum the time intervals $\Delta t_n$ in Eq.~(\ref{tn-2}) to find
\begin{equation}
\label{tn-3}
t_n=
\tau
\left[
1
-
\frac{\tan\tfrac{\theta}{2}}{\tan (n-\tfrac{1}{2})\theta}
\right],
\end{equation}
where $\tau$ is the time interval for the block to reach the
wall when the ball is absent ($\alpha=0$):
\begin{equation}
\label{tau}
\tau={L}/{U}.
\end{equation}
This analytic expression (\ref{tn-3}) for $t_n$ is new.

\begin{figure}[ht]
\begin{center}
\includegraphics[width=80mm]{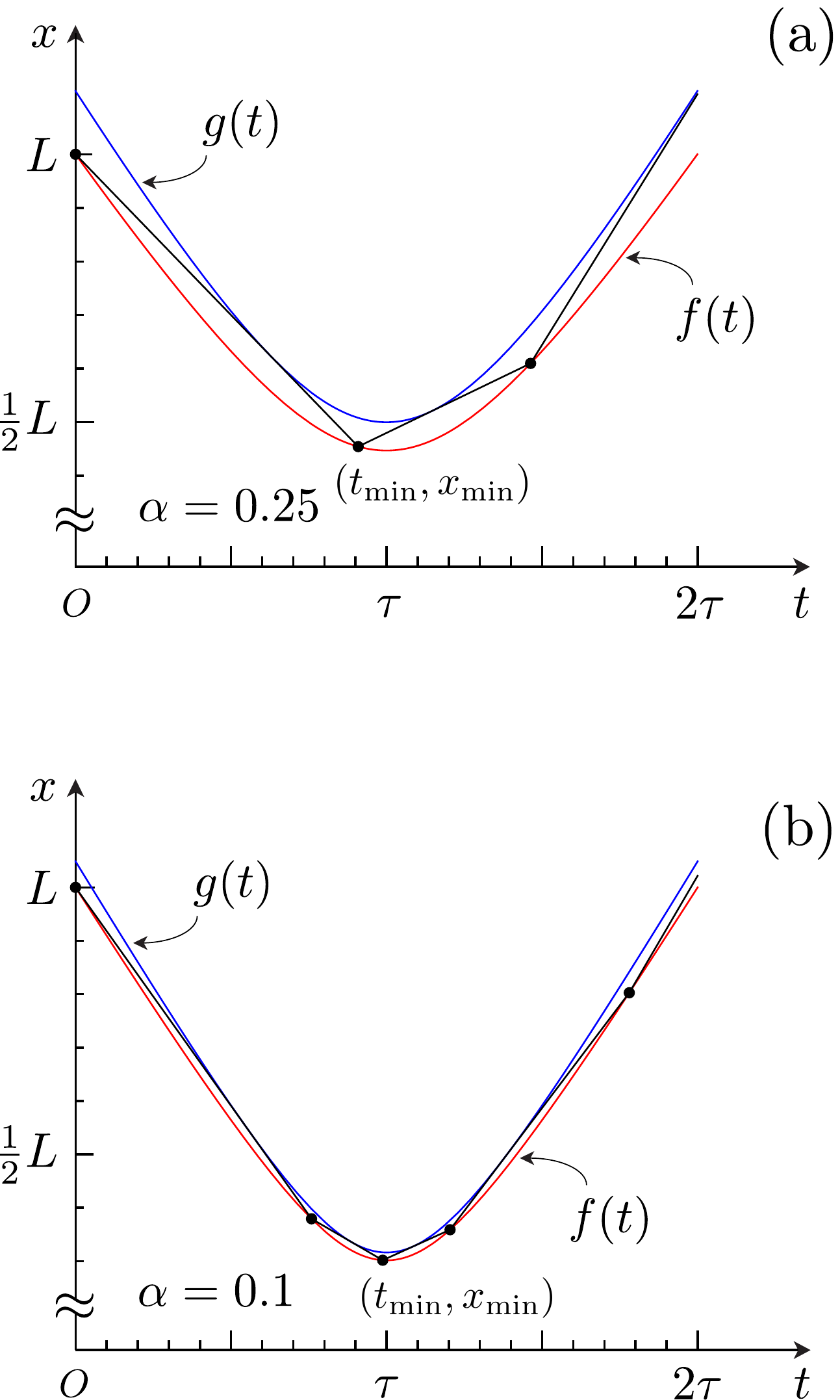}\\
\caption{\label{fig:fg}
(a) The collision points $P_n(t_n,x_n)$ at $\alpha=0.25$, where
$N_1=2$ and $N=3$. The trajectory of the block $x(t)$ consists of
line segments that connect every collision point $P_n$. The lower
bound $f(t)$ passes every collision point $P_n$ and the upper bound
$g(t)$ is tangent to every line segment $\overline{P_nP_{n+1}}$.
The continuum representation $\bar{x}(t)$ of the trajectory is 
the arithmetic average of $g(t)$ and $f(t)$. 
(b) The same as (a) except that $\alpha=0.1$, $N_1=3$,
and $N=5$.
}
\end{center}
\end{figure}

The complete trajectory of the block is
\begin{equation}
\label{x-t}
x(t)=x_n+U_n(t-t_n),\quad t_n\le t\le t_{n+1},
\end{equation}
where $x_n$, $U_n$, and $t_n$ are given in Eqs.~(\ref{xn-3}),
(\ref{sol-vnf}), and (\ref{tn-3}), respectively. 
In Fig.~\ref{fig:fg}, we plot the trajectory $x(t)$ of the
block as a collection of straight line segments.
The vertices represent the collision points $P_n(t_n, x_n)$.
The point on the vertical axis represents the initial collision
point $P_1(0, L)$.
The turning point is
$P_{N_1}(t_{\textrm{min}}, x_{\textrm{min}})$.
When $\alpha = 0.25$ $(0.1)$, we have $N_1=2$ $(3),$ and $N=3$ $(5)$.
The $x$-$t$ plot is convex-downward because every impact on the
the block is along the positive $x$ axis.
\subsection{Computation of $\bm{x_{\rm{min}}}$ and $\bm{t_{\rm{min}}}$}
Next we find the turning point $x_{\textrm{min}}=x_{N_1}$ and 
corresponding time $t_{\textrm{min}}=t_{N_1}$.
According to Eq.~(\ref{NN1}), $(N_1-\tfrac{1}{2})\theta$ is
bounded by 
\begin{equation}
\label{n1range}
\tfrac{1}{2}(\pi-\theta)
\le
(N_1-\tfrac{1}{2})\theta
<\tfrac{1}{2}(\pi+\theta).
\end{equation}
Therefore,
$\cos\frac{\theta}{2}\le \sin(N_1-\tfrac{1}{2})\theta\le 1$. 
Substituting this constraint into Eq.~({\ref{xn-3}}) with
$n=N_1$, we can determine the range of $x_{\rm min}$ as
$L\sin\frac{\theta}{2}\le x_{N_1}\le L\tan\frac{\theta}{2}$.
With the values for $\sin\frac{\theta}{2}$ and 
$\tan\frac{\theta}{2}$ in Table \ref{table:trig}, we obtain
\begin{equation}
\label{xmin-range}
\frac{L\sqrt{\alpha}}{\sqrt{1+\alpha}}
\le
x_{\rm{min}}\le L\sqrt{\alpha}.
\end{equation}
When $\alpha$ is small, $x_{\rm min}\approx L\sqrt{\alpha}$.
From the range in Eq.~(\ref{n1range}) we find that
$\cot\frac{\theta}{2}\le \tan(N_1-\tfrac{1}{2})\theta < \infty$ or
$-\infty<\tan(N_1-\tfrac{1}{2})\theta < -\cot\frac{\theta}{2}$. 
Therefore, considering both cases, we obtain
\begin{equation}
\label{tmin-range}
\tau(1-\alpha)\le t_{\rm{min}}<\tau(1+\alpha),
\end{equation}
where we used $\tan^2\frac{1}{2}\theta=\alpha$. The errors of
$x_{\rm{min}}$ and $t_{\rm{min}}$ are 
$\pm\frac{1}{2}L\sqrt{\alpha}(1-1/\sqrt{1+\alpha})$
and $\pm\alpha \tau$, respectively.

It is interesting to notice that
when $\alpha$ is sufficiently small, 
the time interval $t_{\textrm{min}}$ 
for the block to reach the turning
point is approximately the same as $\tau$, 
which is the time spent for the
block to hit the wall when the ball is missing.
\section{\label{sec:continuum}Continuum limit}
According to Eq.~(\ref{tn-1}), $\Delta t_n$ becomes small when 
$\alpha$ is sufficiently small and $n\approx N_1$, where 
$\sin (n-\frac{1}{2})\theta\approx 1$. This does not require 
that $\alpha\to 0^+$. We call this the \textit{continuum limit}.
In this limit, both $x_n$ and $t_n$ can be treated to be continuous
variables. 

In this section, we investigate the dynamics of the system 
in the continuum limit. As the first step, we express the
collision point $x_n$ in terms of $t_n$. Next, we introduce
a differentiable function $\bar{x}(t)$ that may represent the 
trajectory of the block in the limit. By making use
of the kinetic energy conservation, we
interpret the kinetic energy of the ball as the potential energy
of the block. Expressing the kinetic energy of the ball in terms
of the differentiable function $\bar{x}(t)$,
we compute the force on the block in the continuum limit.
\subsection{Relation between 
$\bm{x_n}$ and $\bm{t_n}$}
We observe that both $x_n$ in Eq.~(\ref{xn-3}) and $t_n$ 
in Eq.~(\ref{tn-3}) depend on $n$ explicitly.
In this section, we find the analytic expression for $x_n$
as a function of $t_n$ by eliminating the explicit $n$
dependence.

From Eqs.~(\ref{xn-3}) and (\ref{tn-3}), we have
\begin{subequations}
\label{tan-n-half}
\begin{eqnarray}
\label{cos-n-half}
x_n\cos(n-\tfrac{1}{2})\theta
&=&
L\cdot\frac{1-t_n/\tau}{\sqrt{1+\alpha}},
\\
\label{sin-n-half}
x_n\sin(n-\tfrac{1}{2})\theta
&=&
L\,\sqrt{\frac{\alpha}{1+\alpha}},
\end{eqnarray}
\end{subequations}
where $\sin\tfrac{\theta}{2}$ and
$\tan\tfrac{\theta}{2}$ are replaced with the values 
listed in Table~\ref{table:trig}. By making use of the
fact that $\sin(n-\tfrac{1}{2})\theta\ge 0$ for all $n\le N$,
we can express $x_n$ as a function of $t_n$:
\begin{equation}
\label{xn-tn-1}
x_n
=\frac{L}{\sqrt{1+\alpha}}
\sqrt{(1-t_n/\tau)^2+\alpha},
\end{equation}
where we used relations in Eq.~(\ref{tan-n-half}) and 
Table~\ref{table:trig}. 

In Appendix~\ref{app-fg}, we have shown that the complete trajectory
$x(t)$ of the block is always bounded by
\begin{equation}
f(t)\le x(t)\le g(t),
\end{equation}
where the differentiable functions $f(t)$ and $g(t)$ are defined by
\begin{subequations}
\begin{eqnarray}
\label{fg-funtions}
f(t) &=& \frac{L}{\sqrt{1+\alpha}}\sqrt{(1-t/\tau)^2+\alpha},
\\
g(t) &=& L\sqrt{(1-t/\tau)^2+\alpha}.
\end{eqnarray}
\end{subequations}
As shown in Fig.~\ref{fig:fg}, the lower bound $f(t)$
passes every collision point $P_n$ and the upper bound $g(t)$ is 
tangent to every line segment $\overline{P_nP_{n+1}}$. 

In the continuum limit, it is convenient to use the arithmetic
average $\bar{x}(t)=\frac{1}{2}[g(t)+f(t)]$ to represent the
trajectory of the block with the uncertainty
$\delta\bar{x}(t)=\frac{1}{2}[g(t)-f(t)]$:
\begin{subequations}
\label{x-Dx-bar}
\begin{eqnarray}
\label{x-bar}
\bar{x}(t)
&=&\mathcal{N}L\sqrt{(1-t/\tau)^2+\alpha},
\\
\label{D-x-bar}
\delta\bar{x}(t)
&=&\Delta L\sqrt{(1-t/\tau)^2+\alpha},
\end{eqnarray}
\end{subequations}
where $\mathcal{N}=\tfrac{1}{2}[1+(1+\alpha)^{-\tfrac{1}{2}}]$
and $\Delta=\tfrac{1}{2}[1-(1+\alpha)^{-\tfrac{1}{2}}]$.
In the limit $\alpha\to 0^+$, all of the curves $\bar{x}(t)$,
$f(t)$, and $g(t)$ collapse into $x(t)$. In addition,
\begin{equation}
\lim_{\alpha\to 0^+}x(t)
=
L|1-t/\tau|,
\end{equation}
which is the case that the ball is absent.

In the continuum limit we can compute the velocity and the
acceleration of the block using $\bar{x}(t)$:
\begin{subequations}
\begin{eqnarray}
\label{Ut}
\dot{\bar{x}}(t) &=&
- \frac{\mathcal{N}\,U(1-t/\tau)}{\sqrt{(1-t/\tau)^2+\alpha}},
\\
\label{acc-t}
\ddot{\bar{x}}(t) &=&
\frac{\alpha\,\mathcal{N} U/\tau}{[(1-t/\tau)^2+\alpha]^{3/2}}.
\end{eqnarray}
\end{subequations}
The values for $\bar{x}$, $\dot{\bar{x}}$, and $\ddot{\bar{x}}$
must be understood as the time-averaged values. The
uncertainties of $\dot{\bar{x}}(t)$ and $\ddot{\bar{x}}(t)$
can be obtained by replacing $\mathcal{N}$ with $\Delta$.
\subsection{Kinetic energy of the block}
Based on Eqs.~(\ref{Ut}) and (\ref{x-bar}), we can
determine the kinetic energy $K$ of the block in terms
of $t$ or $x$ as
\begin{subequations}
\begin{eqnarray}
\label{kinetic-t}
K(t)&=&K_0\frac{\mathcal{N}^2(1-t/\tau)^2}{(1-t/\tau)^2+\alpha},
\\
\label{kinetic-x}
K(x)&=&K_0\mathcal{N}^2\left[
1-\frac{\alpha\mathcal{N}^2L^2}{x^2}
\right],
\end{eqnarray}
\end{subequations}
where $K_0=\tfrac{1}{2}MU^2$ is the initial kinetic energy, 
which can be approximated as $K_0 \approx K(t=-\infty)$.
For a small $\alpha$, the relative error of the kinetic energy is 
$\pm\tfrac{1}{2}\alpha$.
\subsection{Potential energy of the block}
We recall that the total kinetic energy is conserved during the whole 
process. Therefore, the sum of the kinetic energies of the block (${K}$)
and the ball (${\Phi}$)  must be
\begin{equation}
\label{conservation}
K+\Phi=K_0.
\end{equation}
We interpret the kinetic energy $\Phi$ of the ball as the potential
energy of the block. By making use of Eqs.~(\ref{kinetic-t}), 
(\ref{kinetic-x}), and (\ref{conservation}), we find that
\begin{subequations}
\begin{eqnarray}
\label{Phi-t}
\Phi(t)
&=&
K_0\left[1-\frac{\mathcal{N}^2(1-t/\tau)^2}{(1-t/\tau)^2+\alpha}
\right],
\\
\label{Phi-x}
\Phi(x)
&=&
K_0\left[1-\mathcal{N}^2+\frac{\alpha\mathcal{N}^4L^2}
{x^2}\right].
\end{eqnarray}
\end{subequations}
Next we compute the force,
\begin{equation}
\label{F-x}
F(x)=-\tfrac{d\Phi(x)}{dx}= \frac{\alpha \mathcal{N}^{4} MU}{\tau}\cdot 
\frac{L^3}{x^3}.
\end{equation}
The force on the block has the same position dependence as an electric dipole interaction: 
$F(x)\sim 1/x^3$. In addition, $x=x_{\rm min}$ when $t=\tau$.
Therefore, the duration of time to reach $x_{\rm min}$ is
the same as the case when the ball is absent.
As is stated earlier, the derivation of the $1/x^3$ rule originates
from the analytic solution in Eq.~(\ref{x-Dx-bar}), which is in contrast to the approach in
Ref.~\cite{Sinai:1991}.

We can compute $x_{\rm{min}}$ in the continuum limit. Setting
$\Phi(x_{\rm{min}})=K_0$, we find that
\begin{subequations}
\begin{eqnarray}
\label{xmin}
x_{\rm{min}} &=&\mathcal{N} L\sqrt{\alpha},
\\
\label{tmin}
t_{\rm{min}} &=&\tau,
\end{eqnarray}
\end{subequations}
These values are consistent with Eqs.~(\ref{xmin-range}) 
and (\ref{tmin-range}) within errors.
\section{\label{sec:concl}Conclusion}
We have considered the one-dimensional elastic collisions of
a ball and a block against a rigid wall. The initial state
of the ball is $u_0=0$ at $x=L$ and that of the block is
$U_0=-U$. The trajectory $x(t)$ of the block is completely
determined in an analytic form. The analytic expression for
the total number $N$ of collisions between the block and the
ball was also derived. The turning point of the block
is $\approx L\sqrt{\alpha}$ at $t\approx \tau=L/U$,
which is the time interval for the block to hit the wall
when the ball is absent.
Here $\alpha$ is the ratio of the mass of the ball to 
the mass of the block.

In the continuum limit where $\alpha$ is small and $x(t)$ is
near the turning point, the trajectory can be approximated as
a differentiable function $\bar{x}(t)$ in Eq.~(\ref{x-bar}).
Because the total kinetic energy is conserved in this system,
one can think of the kinetic energy of the ball as the
potential energy of the block. Based on this idea, we have
computed the force on the block and found that the force is
proportional to $1/x^3$. It is remarkable that this is a 
unique pure mechanical system that reveals repulsive dipole-like
interaction.
\appendix
\section{Useful Formulas to compute $\bm{t_n}$\label{app-tn}}
The time interval
between $P_{n}$ and $P_{n+1}$
is defined by $\Delta t_n=t_{n+1}-t_n$
which can be expressed
in terms of $U_n$ or $u_n$ as
\begin{equation}
\label{eq-xn}
\Delta t_n 
=\frac{x_{n+1}-x_{n}}{U_n}
=\frac{s_n}{|u_n|},
\end{equation}
where $s_n=x_{n}+x_{n+1}$ is the distance that the ball travels
between $P_n$ and $P_{n+1}$. Assuming $u_n<0$, we find that
\begin{equation}
\label{xn1xn-app}
x_{n+1}=x_{n}(u_n-U_n)/(u_n+U_n).
\end{equation}
Substituting Eq.~(\ref{xn1xn}) into Eq.~(\ref{eq-xn}),
we can express $\Delta t_n$ as
\begin{equation}
\label{tn-rel}
\Delta t_n = -\frac{2x_n}{U_n+u_n}.
\end{equation}
The $n$-th collision time $t_n$ can be computed as
\begin{equation}
\label{delta-tn}
t_n=\sum_{k=1}^{n-1}\Delta t_k.
\end{equation}
\section{Computation of $\bm{U_n}$ and $\bm{u_n}$\label{app-Uun}}
In this section, we solve Eq.~(\ref{Uun-2}),
\begin{equation}
\label{Uun-2-app}
\begin{pmatrix}
U_{n}\\u_{n}
\end{pmatrix}
=A^n
\begin{pmatrix}
U_0\\u_0
\end{pmatrix},
\end{equation}
where the matrix $A$ is defined in Eq.~(\ref{matrix-A}),
\begin{equation}
\label{matrix-A-xxx}
A=\frac{1}{1+\alpha}
\begin{pmatrix}
1-\alpha&-2\alpha\\
2&1-\alpha
\end{pmatrix}.
\end{equation}

We make a transformation diagonalizing $A$ such that
\begin{equation}
\begin{pmatrix}
W_{n}\\w_{n}
\end{pmatrix}=R^{-1}
\begin{pmatrix}
U_{n}\\u_{n}
\end{pmatrix},
\quad
R^{-1}AR=
\begin{pmatrix}
\lambda_1&0\\
0&\lambda_2
\end{pmatrix},
\end{equation}
where $R$ is a $2\times 2$ matrix and $\lambda_i$'s are 
the eigenvalues of the matrix $A$. Then, $U_n$ and $u_n$ are
\begin{equation}\label{sol-vn}
\begin{pmatrix}
U_{n}\\u_{n}
\end{pmatrix}
=R
\begin{pmatrix}
\lambda_1^n W_{0}\\
\lambda_2^n w_{0}
\end{pmatrix},
\quad
\begin{pmatrix}
W_{0}\\
w_{0}
\end{pmatrix}=R^{-1}
\begin{pmatrix}
-U\\
0
\end{pmatrix}.
\end{equation}
The $i$-th column vector $R_i$ of $R$ satisfies the equation,
\begin{equation}
\label{eigen}
AR_i=\lambda_i R_i,
\end{equation}
for $i=1$ and $2$.
The corresponding secular equation 
$\textrm{det}[A-\lambda_i\mathbbm{1}]=0$,
where $\mathbbm{1}$ is the $2\times 2$ identity matrix, is
\begin{equation} 
\alpha(1+\lambda_{i})^2+(1-\lambda_{i})^2=0.
\end{equation}
Solving this equation, we obtain
\begin{subequations}
\label{ans-eigen}
\begin{eqnarray}
\lambda_1&=&
e^{+i\theta},
\quad
R_1=
\begin{pmatrix}
1\\
\displaystyle-{i}/{\sqrt{\alpha}}
\end{pmatrix},
\\
\lambda_2&=&
e^{-i\theta},
\quad
R_2=
\begin{pmatrix}
1\\
\displaystyle +{i}/{\sqrt{\alpha}}
\end{pmatrix},
\end{eqnarray}
\end{subequations}
where the parameter $\theta$ is related to the mass ratio $\alpha$ as
\begin{equation}
\label{theta-alpha-app}
\tan\tfrac{\theta}{2}=\sqrt{\alpha}.
\end{equation}
In Sec.~IV of Ref.~\cite{Whelan:1990}, the authors considered the
collisions of two balls near the ground, which can be compared
to our results by neglecting gravity. The matrix $A$ in 
Eq.~(\ref{matrix-A}) is equivalent to the matrix $M$ in Eq.~(13)
of Ref.~\cite{Whelan:1990}. The elements in the second column
have opposite signs because of the difference in the definition 
of $u_n$. The eigenvalues $\lambda_i$ in Eq.~(\ref{ans-eigen})
and the parameter $\theta$ in Eq.~(\ref{theta-alpha}) are
consistent with those in Eq.~(15) of Ref.~\cite{Whelan:1990}.

Here, we consider only the case $0<\alpha< 1$. Note that 
$\theta=0$ for $\alpha=0$, $0<\theta<\tfrac{\pi}{2}$ 
for $0<\alpha< 1$, and $\theta=\tfrac{\pi}{2}$ for $\alpha=1$.
In the second and third columns of Table~\ref{table:trig},
we list the values for trigonometric functions for 
$\tfrac{1}{2}\theta$ and $\theta$, respectively.
Now we determine $R$ and $R^{-1}$ as
\begin{equation}\label{rnf}
R
=
\begin{pmatrix}
1&1\\
\displaystyle -{i}/{\sqrt{\alpha}}& 
\displaystyle{i}/{\sqrt{\alpha}}
\end{pmatrix},
\quad
R^{-1}=
\frac{1}{2}
\begin{pmatrix}
1&i\sqrt{\alpha}\\
1&-i\sqrt{\alpha}
\end{pmatrix}.
\end{equation}
Substituting Eq.~(\ref{rnf}) into Eq.~(\ref{sol-vn}),
we finally determine the velocities
$U_n$ and $u_n$ as functions of $n$ and $\alpha$:
\begin{equation}
\label{sol-vnf-app}
\begin{pmatrix}
U_{n}\\u_{n}
\end{pmatrix}
=
-\,U
\begin{pmatrix}
\cos n\theta\\
\displaystyle\frac{1}{\sqrt{\alpha}}\sin n\theta
\end{pmatrix}.
\end{equation}
\section{Computation of the time interval $\bm{\Delta t_n}$\label{app-dtn}}
The time interval $\Delta t_n = t_{n+1}-t_n$
can be computed
by substituting $x_n$ in Eq.~(\ref{xn-3}) and
$U_n+u_n$ in Eq.~(\ref{Uun-ans}) into Eq.~(\ref{tn-rel}).
\begin{equation}
\label{tn-1}
\Delta t_n=
\frac{2\tau\sin^2\tfrac{\theta}{2}}
{\sin(n-\tfrac{1}{2})\theta \sin(n+\tfrac{1}{2})\theta },
\end{equation}
where $\tau$ is the time interval for the block to reach the
wall when the ball is absent ($\alpha=0$), that is, $\tau=L/U$.
Substituting $\alpha=(n-\tfrac{1}{2})\theta$ and 
$\beta=(n+\tfrac{1}{2})\theta$ into the following 
trigonometric identity,
\begin{equation}
\label{idcot}
\frac{\sin(\beta-\alpha)}
{\sin\alpha\sin\beta}
=
\cot\alpha
-
\cot\beta,
\end{equation}
we can simplify Eq.~(\ref{tn-1}) as
\begin{equation}
\label{tn-2}
\Delta t_n=
\tau
\left[
\frac{\tan\tfrac{\theta}{2}}{\tan (n-\tfrac{1}{2})\theta}
-
\frac{\tan\tfrac{\theta}{2}}{\tan (n+\tfrac{1}{2})\theta}
\right].
\end{equation}
\section{Determination of the functions $\bm{f(t)}$ and $\bm{g(t)}$
\label{app-fg}}
The complete trajectory of the block $x(t)$ is a set
of line segments that connects consecutive collision
point such that $x(t_n)=x_n$, where
\begin{equation}
\label{xn-tn-1-app}
x_n
=\frac{L}{\sqrt{1+\alpha}}
\sqrt{(1-t_n/\tau)^2+\alpha}.
\end{equation}

We introduce a differentiable function,
\begin{equation}
\label{xA-t}
f(t) = \frac{L}{\sqrt{1+\alpha}}\sqrt{(1-t/\tau)^2+\alpha},
\end{equation}
that passes every collision point $P_n$ and is convex downward.
Therefore, at any time $t\in [0,t_N]$, $f(t) \le x(t)$, where
the equality holds only at $t = t_n$ for $n = 1, \cdots, N$.

The time derivative of Eq.~(\ref{xA-t}) is
\begin{equation}
\label{UA-t}
\dot{f}(t) = -\frac{U}{\sqrt{1+\alpha}}
\frac{1-t/\tau}{\sqrt{(1-t/\tau)^2+\alpha}}.
\end{equation}
By making use of the special values of the trigonometric 
functions in Table~\ref{table:trig}, we find that
\begin{equation}
\label{Uaverage}
\dot{f}(t_n) 
=-U \cos\tfrac{\theta}{2}\cos(n-\tfrac{1}{2})\theta
= \tfrac{1}{2}(U_{n-1}+U_n),
\end{equation}
where we have used $\cos(\alpha+\beta)+\cos(\alpha-\beta)=
2\cos\alpha\cos\beta$ for $\alpha=(n-\tfrac{1}{2})\theta$
and $\beta=\tfrac{1}{2}\theta$.

If there exists a curve $g(t)$ that is tangent to every line segment
$\overline{P_nP_{n+1}}$ at ${t}'_n\in [t_n,t_{n+1}]$ and 
convex downward, then $g(t)$ is an upper bound
of $x(t)$ for all $t\in [0,t_{N}]$. We require
\begin{equation}
\label{g-constraint}
\dot{g}(t'_n) =U_n=-U\cos n\theta,
\end{equation}
and we set $x'_n=g(t'_n)$. According to 
Eq.~(\ref{Uaverage}), $\dot{g}(t'_n)=%
\dot{f}(t_{n+\frac{1}{2}})/\cos\frac{\theta}{2}$.
Therefore, we find that
\begin{subequations}
\begin{eqnarray}
\label{dot-g}
\dot{g}(t)&=&
-\frac{U(1-t/\tau)}{\sqrt{(1-t/\tau)^2+\alpha}},
\\
{t}'_n 
&=&
t_{n+\tfrac{1}{2}}
=\tau(1-\sqrt{\alpha}\cot n\theta),
\\
\label{bar-xn}
{x}'_n
&=&
x(t_{n+\frac{1}{2}}) =
\frac{L\sin\frac{\theta}{2}}{\sin n\theta}.
\end{eqnarray}
\end{subequations}
Note that $t_n\le t'_n\le t_{n+1}$ and
$\dot{g}(t_n) = -U\cos\left(n-\frac{1}{2}\right)\theta$.
Integrating $\dot{g}(t)$ in Eq.~(\ref{dot-g}) over $t$
and imposing the boundary condition (\ref{bar-xn}), we obtain
\begin{equation}
\label{xB-prot}
g(t) = L\sqrt{(1-t/\tau)^2+\alpha}.
\end{equation}
As a result, the trajectory $x(t)$ is bounded by
\begin{equation}
f(t)\le x(t)\le g(t),
\end{equation}
for any $t\in[0,t_N]$. The left and right equalities hold at $t = t_n$
and at $t = t'_n=t_{n+\frac{1}{2}}$, respectively, for all $n$.
In Fig.~\ref{fig:fg}, we show the trajectories of the block $x(t)$
at $\alpha=0.25$ and $\alpha = 0.1$. 
As shown in this figure, the lower bound $f(t)$
passes every collision point $P_n$ and the upper bound $g(t)$ is 
tangent to every line segment $\overline{P_nP_{n+1}}$. 

\begin{acknowledgments}
We thank U-Rae Kim for careful reading of the manuscript
and useful comments.
JL expresses his gratitude to the members of Korea University
Board of Interdisciplinary Communication for an enjoyable discussion
that motivated the work presented here.
The work of JL was supported by Mid-career Research
Program through the NRF grant funded by the MEST (2011-0027559).
\end{acknowledgments}


\newpage
\section*{Tables}
\begin{table}[h]
\begin{center}
\begin{ruledtabular}
\begin{tabular}{cccc} 
$\displaystyle f(x)\,\backslash\,x$ & 
$\displaystyle \tfrac{1}{2}\theta$ & 
$\displaystyle \theta$ & 
$\displaystyle (n-\tfrac{1}{2})\theta$ 
\\[1ex]
\hline
&&&\\[-2.5ex]
$\displaystyle \cos x$ \phantom{xx}&
$\displaystyle \frac{1}{\sqrt{1+\alpha}}$ & 
$\displaystyle \frac{1-\alpha}{1+\alpha}$ &
$\displaystyle 
\frac{1-t_{n}/\tau}{\sqrt{(1-t_{n}/\tau)^2+\alpha}}$ 
\\[3ex]
$\displaystyle \sin x$ \phantom{xx}& 
$\displaystyle \frac{\sqrt{\alpha}}{\sqrt{1+\alpha}}$ & 
$\displaystyle \frac{2\sqrt{\alpha}}{1+\alpha}$ &
$\displaystyle 
\frac{\sqrt{\alpha}}{\sqrt{(1-t_{n}/\tau)^2+\alpha}}$
\\[3ex]
$\displaystyle \tan x$ \phantom{xx}& 
$\displaystyle \sqrt{\alpha}$ & 
$\displaystyle \frac{2\sqrt{\alpha}}{1-\alpha}$ & 
$\displaystyle \frac{\sqrt{\alpha}}{1-t_{n}/\tau}$ 
\end{tabular}
\end{ruledtabular}
\caption{\label{table:trig}
The values for $\cos x$, $\sin x$, and $\tan x$ for
$x=\tfrac{1}{2}\theta$, $\theta$, and $(n-\tfrac{1}{2})\theta$.
The values at $x=n\theta$ and $(n+\tfrac{1}{2})\theta$
can be obtained by using the identities such as
$\cos(x\pm\delta)=\cos x \cos\delta\mp\sin x\sin\delta$ and
$\sin(x\pm\delta)=\sin x \cos\delta\pm\cos x\sin\delta$.
}
\end{center}
\end{table}
\section*{Figures}
\end{document}